\documentclass[12pt]{article}
\usepackage{times}
\usepackage{geometry}
\usepackage[utf8]{inputenc}
\usepackage{enumitem,amssymb}
\usepackage[justify]{ragged2e}
\newlist{thematic}{itemize}{8}
\setlist[thematic]{label=$\square$}
\usepackage{pifont}
\usepackage{graphicx}
 \usepackage[numbers]{natbib}
 \usepackage{wasysym}

 \usepackage{color}
 
\usepackage[usenames,dvipsnames,svgnames,table]{xcolor}
\usepackage[colorlinks=true,
            linkcolor=blue,
            urlcolor=blue,
            citecolor=blue]{hyperref}

\begin{document}
\setlength{\textfloatsep}{8pt plus 1.0pt minus 2.0pt}

\raggedright
\huge
Astro2020 Science White Paper \linebreak

The L/T Transition \linebreak
\normalsize

\noindent \textbf{Thematic Areas:} \hspace*{55pt} $\XBox$ Planetary Systems \hspace*{14pt} $\Square$ Star and Planet Formation \hspace*{20pt}\linebreak
$\Square$ Formation and Evolution of Compact Objects \hspace*{31pt} $\Square$ Cosmology and Fundamental Physics \linebreak
  $\XBox$  Stars and Stellar Evolution \hspace*{1pt} $\Square$ Resolved Stellar Populations and their Environments \hspace*{40pt} \linebreak
  $\Square$    Galaxy Evolution   \hspace*{45pt} $\Square$             Multi-Messenger Astronomy and Astrophysics \hspace*{65pt} \linebreak
  
\textbf{Principal Author:}

Name:	Johanna M. Vos
 \linebreak						
Institution:  American Museum of Natural History
 \linebreak
Email: jvos@amnh.org
 \linebreak
Phone:  +1 (212) 769-5996
 \linebreak
 
\textbf{Co-authors:}  \\
Katelyn Allers ~(Bucknell University)\\
Daniel Apai ~(University of Arizona)\\
Beth  Biller ~(University of Edinburgh)\\
Adam J. Burgasser (University of California San Diego)\\
Jacqueline Faherty (American Museum of Natural History)\\
Jonathan Gagn\'e (Universit\'e de Montr\'eal) \\
Christiane Helling (University of St Andrews)\\
Caroline Morley (University of Texas)\\
Jacqueline Radigan (Utah Valley University)\\
Adam Showman (University of Arizona)\\
Xianyu Tan (University of Arizona) \\
Pascal Tremblin (Universit\'e Paris-Saclay)
\linebreak

\begin{justify}
\textbf{Abstract:} \\
 The L/T transition is an important evolutionary phase in brown dwarf atmospheres, providing us with a unique opportunity to explore the effects of clouds, convection, winds, gravity and metallicity across a very narrow temperature range. Understanding these physical processes is critical for understanding ultracool atmospheres. In the next decade, we will answer three key questions regarding L/T transition atmospheres:
     \setlist{nolistsep}
 \begin{enumerate}
 \itemsep0em 
\item What is the physical mechanism behind the L/T transition?
\item What is the spatial extent of atmospheric structures at the L/T transition?
\item How do gravity and metallicity affect the L/T transition?
 \end{enumerate}
The theory and methods developed for brown dwarfs will be used in the 2030s and beyond for solar-system age giant exoplanets and eventually habitable zone earth analogues. Developing these techniques now are crucial.

\pagebreak


\subsection*{Background}


The directly imaged exoplanets observed to date share their temperatures, ages and surface gravities with L and T type brown dwarfs \citep{Kirkpatrick2005}. In the absence of a bright host star, brown dwarfs are much easier to observe in detail and act as powerful exoplanet analogs.
Brown dwarf atmospheres play an influential role on their evolution and observed properties, regulating their radiative cooling and imprinting the signatures of gases and condensates onto their observed spectra.
Thus, the key to understanding the spectra and evolution of brown dwarfs lies in understanding their atmospheric processes.
The L/T transition is an extremely important evolutionary phase in brown dwarf atmospheres, providing us with a unique opportunity to explore the effects of clouds, convection, winds, gravity and metallicity across a very narrow temperature range. Understanding these physical processes is critical for understanding ultracool atmospheres. \smallskip

\begin{figure}[tb]
 \begin{center}
 \includegraphics[scale=0.6]{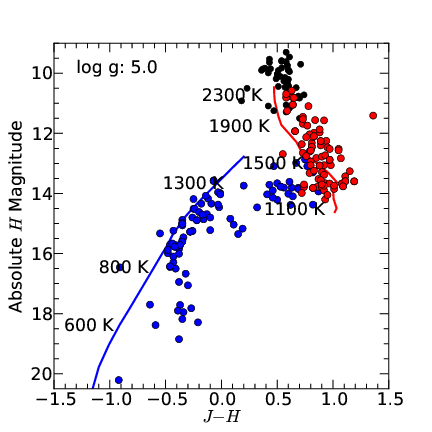}
\end{center}
\caption{Color-magnitude diagram showing L and T spectral type brown dwarfs with measured parallaxes from \citep{Dupuy2012}. Red lines show cloudy models and blue lines show cloud-free models. Figure from \citep{Morley2012}.}
 \label{fig:Morley2012}
\end{figure}

\noindent L dwarf temperatures ($<2200$K) and pressures favor the formation of atmospheric condensates composed of magnesium silicates and iron, collectively referred to as dust \citep{Tsuji1999, Allard2001}. These dust clouds are generally thought to shape the observed properties of L-type brown dwarfs in the IR \citep{Helling2014, Marley2015}. The color-magnitude diagram in Figure \ref{fig:Morley2012} shows the color evolution of a brown dwarf as it cools to later spectral types. Dust clouds are a major source of opacity in the near-IR, which causes L dwarfs to become increasingly redder in the near-IR as the condensate clouds build up with later spectral type. 
As a brown dwarf continues to cool, it undergoes a significant transformation in its observed properties at temperatures $<1400$K. 
Within a very narrow range of effective temperatures ($\sim200$K), the near-IR colors of brown dwarfs shift dramatically bluer and a brightening in the $J$-band is observed for early T dwarfs (Figure \ref{fig:Morley2012}). Cloud models \citep[e.g.][]{Marley2015, Charnay2018} have shown that this sharp transition can be explained by the disappearance of dust clouds as they sink below the photosphere, however the physical mechanism of this transition has remained elusive.
Recently,  models have been proposed that explain the spectra and colours of brown dwarfs as a result of diabatic convection in the atmosphere, triggered by the instability of carbon chemistry in brown dwarf atmospheres\citep{Tremblin2019}.  \smallskip

\begin{figure}[tb]
 \begin{center}
 \includegraphics[scale=0.36]{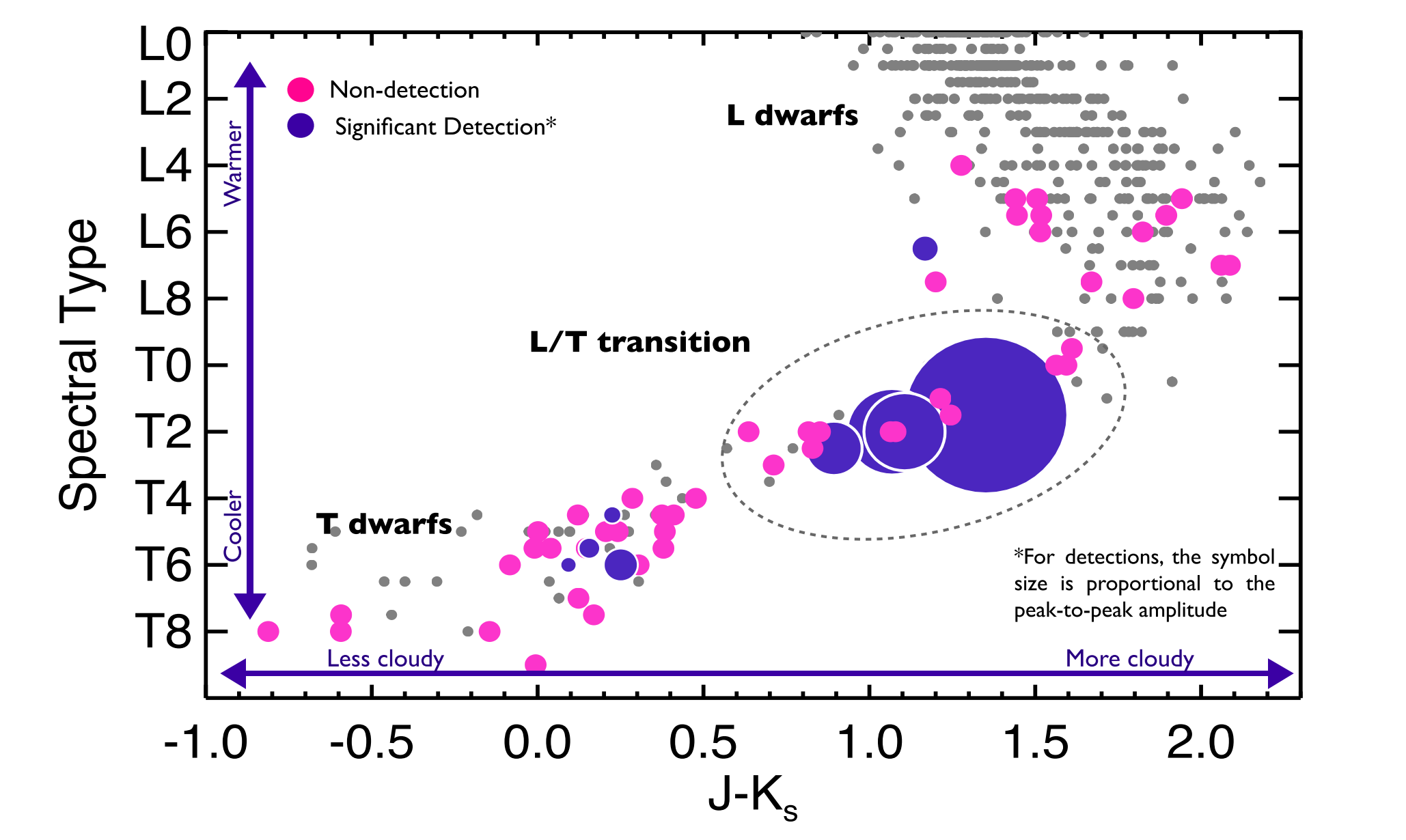}
\end{center}
\caption{\citep{Radigan2014} carried out a ground-based $J$-band survey for variability in L and T type brown dwarfs. An increase in variability amplitude and variability occurrence rate was observed at the L/T transition, shown by the dotted circle.}
 \label{fig:Radigan2014}
\end{figure}

\noindent The combination of rapid rotation and atmospheric inhomogeneities has long motivated searches for rotationally modulated variability. The past decade has seen rapid advancement in variability studies of brown dwarfs. Initial detections of high-amplitude variability in a sample of L/T transition brown dwarfs lent support to patchy cloud models \citep{Artigau2009, Radigan2012, Gillon2013}. These studies were followed by large surveys, which found ubiquitous variability across the entire L--T spectral sequence \citep{Metchev2015a}, with an increase in amplitude and occurrence rate across the L/T transition \citep[Figure \ref{fig:Radigan2014};][]{Radigan2014}.
A number of variable objects have been studied in detail with spectroscopic monitoring \citep{Apai2013, Buenzli2015, Yang2016, Schlawin2017}. Combining the observations with cloud models indicates the presence of multiple cloud layers affecting the emergent flux. 
{Since variability directly probes the presence of atmospheric inhomogeneities as they rotate in and out of view, observations of brown dwarf variability can reveal the detailed mechanisms driving the L/T transition.}

\subsection*{1. What is the physical mechanism behind the L/T transition?}


 Current state-of-the-art 1D cloud models incorporate non-uniform clouds across the L/T transition in order to match the observed colors and spectra of these objects \citep{Morley2014, Charnay2018}, however no consensus has been reached on the physical mechanism for the transition from cloudy to cloud-free atmospheres. 
 \citep{Tan2018} have presented a promising mechanism for cloud dispersal at the L/T transition, finding that clouds with larger particle sizes dissipate more easily than clouds with smaller particles sizes. Early modelling efforts found that L dwarfs are likely dominated by sub-micron particles, T dwarfs with clouds of larger particles sizes \citep[e.g.][]{Burningham2017, Saumon2008}. 
 If the change from L to T spectral types is accompanied by a change from small to large particles, then cloud fragmentation at the L/T transition may naturally result. This potential mechanism requires further study through developing more advanced cloud microphysics models \citep{Helling2008}. \smallskip

\noindent 
The possibility that the spectra and colors of brown dwarfs may be explained via convection mechanisms rather than condensate clouds raises another challenge \citep{Tremblin2015, Tremblin2016, Tremblin2017}. 
Brown dwarf atmospheres are most likely affected by both effects, and there is a real need for cloud models to incorporate these additional atmospheric processes and vice versa.  It is critical that theorists have access to the necessary funding to do this work, which will reveal both the nature of the L/T transition and the evolution of planetary atmospheres. It is also important that the convection models be tested on the large body of spectroscopic variability data \citep{Apai2013,Buenzli2015} that have been interpreted as evidence for the presence and importance of condensate clouds in shaping observed properties. Developing these models will reveal the effects of diabatic convection on L/T transition atmospheres. \smallskip
 
\noindent In order to validate atmospheric models, high-quality observations of brown dwarf variability is crucial.
 In the next decade we have the opportunity to characterise the large number of known variable brown dwarfs \citep[e.g.][]{Artigau2009,Gillon2013,Radigan2014, Metchev2015a} to an unprecedented extent, particularly in the mid-IR. 
 Since the end of the Spitzer cryogenic mission in 2009, wavelengths beyond $5~\mu $m have been inaccessible for variability monitoring observations since ground-based facilities suffer from very large sky background at these long wavelengths. These wavelengths provide crucial information on the atmospheric properties of L/T transition brown dwarfs, particularly the $10~\mu $m silicate scattering feature \citep{Cushing2006}, which will enable us to directly probe the presence and distribution of silicate grains in the atmosphere. \citep{Tremblin2017} find that mid-IR spectra in the $3-7~\mu $m region will directly determine the presence of clouds and their importance on the emergent flux. Spectroscopic variability monitoring at these wavelengths also deeply probe the vertical cloud structures in the $5~\mu$m window, allowing us to probe 3D atmospheres. Variability monitoring observations with JWST/MIRI will be crucial for this work.

 \subsection*{2. What is the spatial extent of atmospheric structures at the L/T transition?}

 While 1D models can model the detailed cloud physics that are crucial at the L/T transition, they assume an atmosphere in its steady state. Since the last Decadal Survey, we have found that many variable brown dwarfs exhibit significant evolution in their light curves \citep[e.g][]{Karalidi2016, Apai2017a}, indicative of cloud dynamical evolution. A key question is whether brown dwarfs exhibit banded or isotropic structures, which would indicate rotationally or convectively dominated atmospheric circulation \citep{Showman2013}.
Further development of 3D GCMs will help to answer this question and to reveal the mechanisms driving the long-term variability of L/T transition brown dwarfs.\smallskip
 
 
 \noindent High-quality brown dwarf variability monitoring observations over long periods will be critical for constraining 3D GCM models. Variability observations to date have generally been constrained to continuous monitoring over $\sim2-3$ rotation periods. Even with these relatively short durations, significant light curve evolution has been observed \citep{Artigau2009, Yang2016}. Recently, \citep{Apai2017a} presented the longest continuous Spitzer observations of a sample of three L--T type brown dwarfs, finding light curve evolution in each case. The long-term behaviour of the light curves is best explained by the beating pattern of planetary-like jets with different velocities, a feature observed in our own Solar System on Neptune. Long-term monitoring observations can constrain the size, number and velocities of cloud bands. These must be obtained over longer duration observations than those obtained to date (weeks--months), so that we can fully sample the complex structure and  cloud evolution. As discussed in the white paper by \citep{Apai2019}, an IR space-based telescope or network of ground-based telescopes capable of long-term monitoring of brown dwarfs can provide these data. These observations will guide the 3D GCM models over the next decade. \smallskip

\begin{figure}[tb]
 \begin{center}
 \includegraphics[scale=0.32]{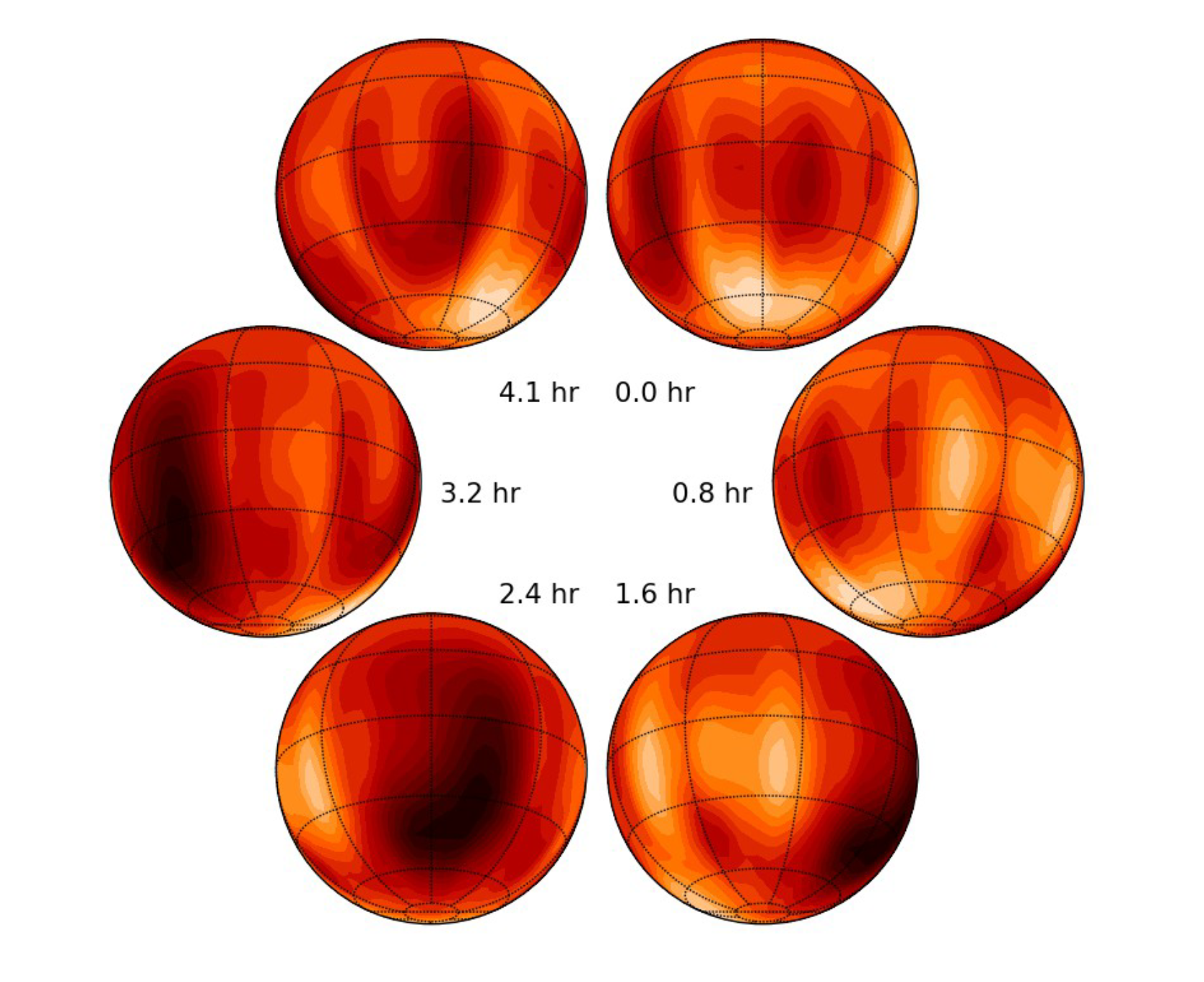}
\end{center}
\caption{Doppler map of Luhman 16B showing the presence of atmospheric inhomogeneities. Doppler maps of a larger sample of L/T brown dwarfs will be necessary to constrain 3D global circulation models. Figure from \citep{Crossfield2014}.}
 \label{fig:Crossfield2014}
\end{figure}

\noindent Spatial mapping is another essential technique for constraining GCMs. Since the last Decadal Survey techniques have been developed to probe the shape and spatial extent of atmospheric features in brown dwarfs atmospheres. The long-term monitoring observations mentioned above are the ideal dataset for further developing the spatial mapping codes developed by \citep{Karalidi2015, Karalidi2016, Apai2017a}. These spatial mapping codes map the surface features (spots or bands) of the atmosphere per observational wavelength using the observed rotational light curves, and will be instrumental in examining the evolution of the atmospheric features themselves once we have obtained the necessary long-term observations. 
\noindent Doppler imaging is a powerful and more direct technique that can map surface features, using the varying Doppler shifts across the face of a rotating object to map inhomogeneous surfaces \citep{Vogt1987}. \citep{Crossfield2014} reported the first (and so far only) Doppler map of a brown dwarf, Luhman 16B (Figure \ref{fig:Crossfield2014}), which depicts a bright near-polar region and a darker mid-latitude area, consistent with large-scale cloud inhomogeneities.  With current instruments, this technique can only be applied to the brightest few brown dwarfs. With next-generation 30 m telescopes, Doppler imaging will become feasible for dozens of brown dwarfs and for a handful of the brightest directly-imaged exoplanets \citep{Crossfield2014b}. Doppler imaging of a large sample of brown dwarfs will reveal whether brown dwarfs have banded structures (like Jupiter) or whether they are dominated by more isotropic turbulence, and ascertain the dominant length scales of atmospheric structures. The importance of Doppler imaging is also discussed in white papers \citep{Apai2019} and \citep{Burgasser2019}. 

\subsection*{3. How do gravity and metallicity affect the L/T transition?}

\textbf{Gravity:~~}In the last decade, low-gravity brown dwarfs have emerged as a particularly important subset of brown dwarfs, due to their physical similarities with directly-imaged exoplanets. 
Low-gravity L dwarfs follow a distinct sequence in the color-magnitude diagram, appearing consistently redder and less luminous in the near-IR than their higher-gravity field brown dwarf counterparts \citep{Faherty2016}. To explain these differences, models have invoked thicker condensate clouds and non-equilibrium chemistry, which arise as a result of their low-gravity \citep{Marley2012}. Indeed, recent variability studies of low-gravity L dwarfs show different variability properties compared to field objects, suggesting different cloud properties for low-gravity atmospheres \citep{Biller2015, Vos2019}. \smallskip

\noindent There appears to be a delay in the temperature at which an L dwarf enters the L/T transition for directly-imaged planets and low-gravity brown dwarfs \citep{Metchev2006, Macintosh2015, Faherty2016}, however the few low-gravity objects with spectral types $>$L9 currently known prohibits us from accurately modelling the effect of surface gravity at the L/T transition. Currently, we have a precise temperature constraint for two L/T transition objects with a known age. The T2 AB Doradus member 2MASS1324+63 and the T2.5 Carina-Near member SIMP0136 have temperatures cooler than field L/T transition brown dwarfs \citep{Gagne2017, Gagne2018a}.
In order to understand the gravity effects at play, it is critical that we increase our current sample of young L/T transition objects. In the last decade, a large sample of low-gravity L dwarfs were discovered by identifying members of young moving groups and by pinpointing signatures of low-gravity in their spectra \citep{Cruz2009, Allers2013}. 
In the next decade we will use these same techniques on the cooler, fainter objects at the L/T transition and beyond. High-quality spectra obtained using $8$ m telescopes will allow us to identify the signatures of youth, and constrain atmospheric models at these temperatures. As discussed in the white paper by \citep{Faherty2019}, the next decade will reveal an unprecedented library of very low-mass objects that are essential for advancing knowledge on the atmospheric properties of the directly-imaged exoplanets. \smallskip


\noindent
\textbf{Metallicity:~}
As condensates are inherently metal-rich, atmospheric abundances plays a role in determining the chemistry and spectral evolution across the L/T transition. Metal-poor halo L subdwarfs \citep{Burgasser2003, Zhang2017} have been shown to exhibit bluer near-infrared spectral energy distributions, indicating less absorption by condensates clouds in the 1 $\mu$m region as compared to solar-metallicity L dwarfs, as well as enhanced collision-induced H$_2$ absorption. Unlike M subdwarfs, double-metal oxides are enhanced in the optical spectra of these objects, suggesting condensate formation is suppressed or delayed at low metallicities \citep{Gizis2006, Burgasser2007}.  This is likely to have an impact on the metal-poor L/T transition, although no very metal-poor T subdwarfs ([M/H] $\lesssim$ -1) have been uncovered to date \citep{Zhang2018}. Unusually blue L dwarfs, which appear to be modestly metal-poor ([M/H] $\gtrsim$ -0.5), are well-matched to cloud models with thin clouds or enhanced condensate rainout \citep{Burgasser2008, Cushing2010}; while the photospheres of blue T dwarfs, like regular T dwarfs, appear largely condensate-free.  One the other end of the metallicity spectrum, \citep{Looper2008} found that unusually red L dwarfs could be equally matched to both low surface gravity and superstar metallicity models, each case producing red spectral energy distributions consistent with thick condensate clouds. These results indicate a complicated interplay between surface gravity, metallicity and cloud content that may shape the L/T transition in different ways. The specific role of metallicity could be clarified by building and analyzing a metallicity-calibrated sample of benchmark L and T dwarfs; e.g., companions to FGK stars with precision metallicities \citep[e.g. Wolf 1130B;][]{Mace2013} and globular cluster members.  Deeper infrared surveys will also be needed to uncover a sufficient sample of halo T subdwarfs to trace the L/T transition at subsolar metallicities.

\end{justify}

{\footnotesize{
\bibliographystyle{apj} 
\bibliography{Full} }}
\end{document}